\documentclass[11pt]{article}
\usepackage{amsmath}
\usepackage{latexsym}
\usepackage{amssymb}
\usepackage{graphicx}

\begin{document}

\title{The Universe is a Strange Place\footnote{Keynote talk at SpacePart03,
December 2003.}}
\author{F. Wilczek\footnote{Email: wilczek@mit.edu}\\
\\
{\small\itshape Center for Theoretical Physics}
   \\%[-1ex]
{\small\itshape Department of Physics}\\%[-1ex]
{\small\itshape Massachusetts Institute of
Technology} \\
{\small\itshape  Cambridge, Massachusetts 02139}}

\date{\small MIT-CTP-3465}

\maketitle

\pagestyle{myheadings} \markboth{F. Wilczek}{The Universe is a Strange Place}
\thispagestyle{empty}

\begin{abstract}
This is a broad and in places unconventional overview of the strengths and
shortcomings of our standard models of fundamental physics and of cosmology.
The emphasis is on ideas that have accessible empirical consequences.
It becomes
clear that the frontiers of these subjects share much ground in common.
\end{abstract}

\section{Standard Models}

Our knowledge of the physical world has reached a most remarkable state.
We have established economical ``standard models" both for cosmology
and for fundamental physics, which provide the conceptual foundation
for describing a vast variety of phenomena in terms of a small number of
input parameters.  No existing observations are in conflict with
these standard models.

This achievement allows us to pose, and have genuine prospects to
answer, new questions
of great depth and ambition.  Indeed, the standard models themselves,
through their weirdness and
esthetic failings, suggest several such questions.

A good way to get a critical perspective on these standard models is to review
the external inputs they require to make them go.  (There is a nice
word for such input parameters: ``exogenous",
born on the outside. ``Endogenous" parameters, by contrast, are
explained from within.)
This exercise also exposes interesting things about the nature of the models.

\subsection{Fundamental Physics}

What is usually called the standard model of particle physics is actually
a rather arbitrary truncation of our knowledge of fundamental physics,
and a hybrid to boot. In is more accurate and more informative to speak of
two beautiful theories and one rather ramshackle working model:
the gauge theory, the gravity theory, and the flavor/Higgs model.

The gauge theory is depicted in Figure 1. It is based on the local
(Yang-Mills) symmetry $SU(3)\times SU(2) \times U(1)$.   The fermions 
in each family fall
into 5 separate irreducible representations of this symmetry. They are
assigned, on phenomenological grounds, the funny $U(1)$ hypercharges
displayed there as subscripts.  Of course, notoriously, the whole family
structure is triplicated. One also needs a single $SU(3)$ singlet, $SU(2)$
doublet scalar ``Higgs" field with hypercharge $-\frac{1}{2}$.
\begin{figure}[t]
\begin{center}
\framebox{
\includegraphics{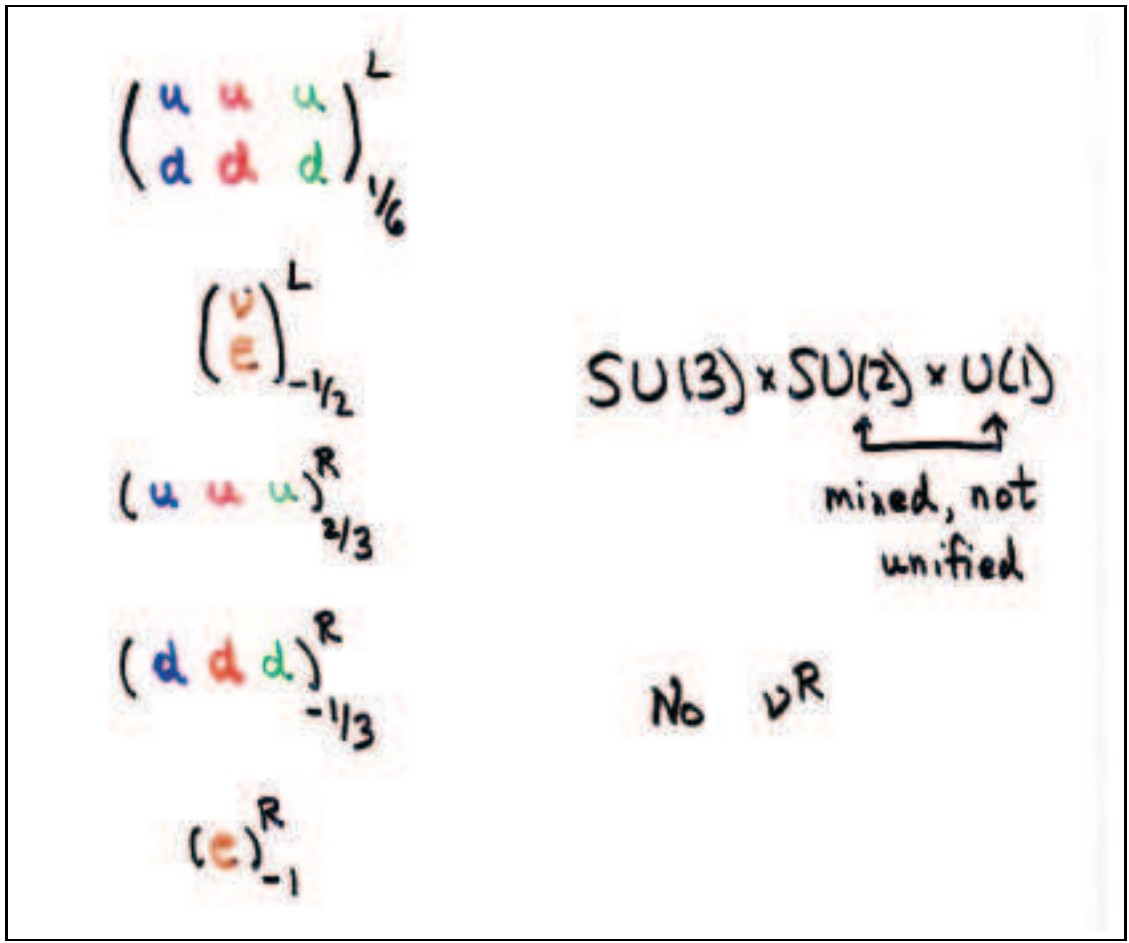}
}
\caption{Specification of the gauge sector. See text.}
\end{center}
\vspace{-4pt}
\end{figure}

Taken together, gauge symmetry and renormalizability greatly restrict
the number of independent couplings that are allowed.
Putting aside for a moment Yukawa couplings of fermions and Higgs field,
there are just three continuous parameters, namely the universal interaction
strengths of the different gauge fields.
(There is a subtlety here regarding the $U(1)$ charges. Since
implementing $U(1)$ gauge symmetry does not
automatically quantize charge, the hypercharge assignments
might appear to involve many continuous parameters, which on phenomenological
grounds we must choose to have extremely special values. And
that is true, classically. But consistency of the quantum theory requires
cancellation of anomalies. This requirement greatly constrains the possible
hypercharge assignments, bringing us down essentially to the ones we adopt.)

I should emphasize that gauge symmetry and renormalizability are
deeply tied up with the consistency and existence of quantum field
theories involving vector mesons.
Taking a little poetic license, we could say
that they are not independent assumptions at all, but rather
consequences of special relativity and quantum mechanics.

General relativity manifestly provides a beautiful, conceptually driven
theory of gravity. It has scored many triumphs, both qualitative
(big bang cosmology, black hole physics) and quantitative (precession
of Mercury,
binary pulsar). The low-energy effective theory of gravity and the
other interactions is defined algorithmically by the minimal coupling
prescription, or equivalently by restricting to low-dimension operators.
In this context, ``low" means compared to the Planck energy scale, so this
effective theory is very effective indeed. As in the gauge sector,
symmetry -- here, general covariance -- greatly constrains the
possible couplings,
bringing us down to just two relevant parameters.
Almost all the observed phenomena of gravity are described
using only one of these parameters, namely Newton's gravitational constant.
We are just now coming to accept that the other parameter, the
value of the cosmological term, plays an important role in describing
late-time cosmology.

This impressive effective field theory of gravity is perfectly
quantum-mechanical.
It supports, for example, the existence of gravitons as the
particulate form of gravity waves.
There are major unsolved problems in gravity, to be sure, a few of which
I'll discuss below, but they shouldn't be overblown or made to seem mystical.

The third component of the standard model consists, one might say, of the
potential energy terms.  They are the terms that don't arise from gauge or
space-time covariant derivatives.  (Note that field strengths and
curvatures are commutators of covariant derivatives.)  All these
terms involve the Higgs field, in one way or another. They include
the Higgs field
mass and its self-coupling, and the Yukawa couplings.   We know of no deep
principle, comparable to gauge symmetry or general covariance, which
constrains the values of these couplings tightly. For that reason, it is
in this sector where continuous parameters proliferate, into the dozens.
Basically, we introduce each observed mass and weak mixing angle as
an independent input, which must be determined empirically.  The
phenomenology is not entirely out of control: the general framework
(local relativistic quantum field theory, gauge symmetry, and
renormalizability) has significant consequences, and even this part of
the standard model makes many non-trivial predictions and is highly
over-constrained. In particular, the Cabibbo-Kobayashi-Maskawa
(CKM) parameterization of weak currents and CP violation has, so far, survived
close new scrutiny at the B-factories intact.

Neutrino masses and mixings can be accommodated along similar lines, if
we expand the framework slightly. The simplest possibility is to allow for
minimally non-renormalizable (mass dimension 5) ``ultra-Yukawa"
terms. These terms involve two powers of the scalar Higgs field.
To accommodate the observed neutrino masses and mixings, they must occur with
very small coefficients.

\subsection{Cosmology}

The emerging ``standard model of cosmology" is also something of a hybrid.
One part of it is simply a concrete parameterization of the equation of
state to insert into the framework of general relativistic models of
a spatially
uniform expanding Universe (Friedmann-Robertson-Walker model);
the other is a very specific hypothesis about the primordial fluctuations from
uniformity.

Corresponding to the first part, one set of exogenous parameters in the
standard model of cosmology specifies a few average properties of matter,
taken over large spatial volumes. These are the densities of ordinary
matter (i.e., of baryons), of dark matter, and of dark energy.

We know quite a lot about ordinary matter, of course, and we can detect
it at great distances by several methods. It contributes about 3\% of
the total density.

Concerning dark (actually, transparent) matter we know much less.  It
has been ``seen"
only indirectly, through the influence of its gravity on the motion
of visible matter.
We observe that dark matter exerts very little pressure, and
that it contributes about 30\% of the total density.

Finally dark (actually, transparent) energy contributes about 67\% of the
total density.  It has a large {\it negative\/} pressure. From the point
of view of fundamental physics this dark energy is quite mysterious and
disturbing, as I'll elaborate shortly below.

Given the constraint of spatial flatness, these three densities are
not independent. They must add up to a critical density that
depends only the strength of gravity and the rate of expansion of the universe.

Fortunately, our near-total ignorance concerning the nature of most
of the mass of the Universe does not bar us from modeling its evolution.
That's because the dominant interaction on large scales is gravity, and
according to general relativity gravity does not care about details.
According to general relativity, only total energy-momentum counts -- or
equivalently, for uniform matter, total density and pressure.

Assuming these values for the relative densities, and that the geometry
of space is flat -- and still assuming uniformity -- we can use the equations
of general relativity to extrapolate the present expansion of the Universe
back to earlier times. This procedure defines the standard Big Bang scenario.
It successfully predicts several things that would otherwise be very
difficult to understand, including the red shift of distant galaxies,
the existence of the microwave background radiation, and the relative abundance
of light nuclear isotopes.  It is also internally consistent, and even
self-validating, in that the microwave background is observed to be uniform to
high accuracy, namely to a few parts in $10^5$.

The other exogenous parameter in the standard model of cosmology
concerns the small departures from uniformity in the early Universe.
The seeds grow by gravitational instability, with over-dense regions
attracting more matter, thus increasing their density contrast with time.
This process plausibly could, starting from very small seeds, eventually
trigger the formation of galaxies, stars, and other structures
we observe today.  {\it A priori\/} one might consider all kinds
of assumptions about
the initial fluctuations, and over the years many hypotheses
have been proposed.  But recent observations, especially the recent,
gorgeous WMAP
measurements of microwave background anisotropies, favor what
in many ways is the simplest possible guess, the so-called Harrison-Zeldovich
spectrum. In this set-up the fluctuations are assumed to be strongly
random -- uncorrelated and Gaussian with a scale invariant spectrum at
horizon entry, to be precise -- and to affect both ordinary and dark
matter equally (adiabatic fluctuations).  Given these strong
assumptions just one
parameter, the overall amplitude of fluctuations, defines the statistical
distribution completely. With the appropriate value for this amplitude, and
the relative density parameters I mentioned before, this standard
model cosmological model fits the WMAP data and other measures of large-scale
structure remarkably well.

\section{From Answers to Questions: Fundamental Physics}

The structure of the gauge sector of the standard model gives
powerful suggestions for its further development. The product
structure $SU(3)\times SU(2) \times U(1)$, the reducibility of the fermion
representation, and the peculiar values of the hypercharge
assignments all suggest the possibility of a larger symmetry,
that would encompass the
three factors, unite the representations, and fix the hypercharges.
The devil is in the details, and it is not at all automatic that the observed,
complex pattern of matter will fit neatly into a simple mathematical structure.
But, to a remarkable extent, it does.  The smallest simple group into
which $SU(3)\times SU(2) \times U(1)$ could possibly fit, that is $SU(5)$,
fits all the fermions of a single family into two representations
($\bf{10} +\bar{\bf 5}$), and the hypercharges click into place. A
larger symmetry
group, $SO(10)$, fits these and one additional $SU(3)\times SU(2) \times
U(1)$ singlet particle into a single representation, the spinor $\bf{16}$.
The additional particle is actually quite welcome.  It has the quantum
numbers of a right-handed neutrino, and it plays a crucial role in
the attractive ``seesaw" model of neutrino masses, of which more below.

This unification of quantum numbers, though attractive, remains
purely formal until it is embedded in a physical model.  That
requires realizing the enhanced symmetry in a local gauge theory. But
nonabelian
gauge symmetry requires universality: it requires that the relative
strengths of the different couplings must be equal, which is not what
is
observed.

Fortunately, there is a compelling way to save the situation. If the higher
symmetry is broken at a large energy scale (equivalently, a small
distance scale),
then we observe interactions at smaller energies (larger
distances) whose intrinsic strength has been affected by the physics of
vacuum polarization.  The running of couplings is an effect that can be
calculated rather precisely, in favorable cases (basically, for weak coupling),
given a definite hypothesis about the particle spectrum. In this way
we can test, quantitatively, the idea that the observed couplings derive
from a single unified value.

Results from these calculations are quite remarkable and encouraging.
If we include vacuum polarization from the particles we know about in
the minimal standard model, we find approximate unification. If we include
vacuum polarization from the particles needed to expand the standard
model to include supersymmetry, softly broken at the Tev scale, we find
accurate unification. The unification occurs at a very large energy scale, of
order $10^{16}$ Gev.  This success is robust against small changes in the
SUSY breaking scale, and is not adversely affected by incorporation of
additional particle multiplets, so long as they form complete representations
of $SU(5)$.

On the other hand, many proposals for physics beyond the standard model
at the Tev scale (Technicolor models, large extra dimension scenarios, most
brane-world scenarios) corrupt the foundations of the unification of
couplings calculation, and would render its success accidental. For me,
this greatly diminishes the credibility of such proposals.

Low-energy supersymmetry is desirable on several other grounds, as well.
The most important has to do with the ``black sheep" of the standard model,
the scalar Higgs doublet. In the absence of supersymmetry radiative
corrections to the vacuum expectation value of the Higgs particle diverge,
and one must fix its value (which, of course, sets the scale for electroweak
symmetry breaking) by hand, as a renormalized parameter. That leaves it
mysterious why the empirical value is so much smaller than unification
scales.

Upon more detailed consideration the question takes shape and
sharpens considerably. Enhanced unification symmetry requires that the
Higgs doublet should have partners, to fill out a complete representation.
However these partners have the quantum numbers to mediate proton
decay, and so if they exist at all their masses must be very large, of
order the unification scale $10^{16}$ Gev.  This reinforces the idea
that such a
large mass is what is ``natural" for a scalar field, and that the light doublet
we invoke in the standard model requires some special justification. It
would be facile to claim that low-energy supersymmetry by itself
cleanly solves these problems, but it does provide powerful
theoretical tools for
addressing them.

The fact that an enormous new mass scale for unification is indicated
by these calculations is profound.  This enormous mass scale is
inferred entirely from low-energy data. The disparity of scales
arises from the slow
(logarithmic) running of inverse couplings, which implies that
modest differences in observed couplings must be made up by a long interval of
running.  The appearance of a very large mass scale is welcome on
several grounds.

\begin{itemize}
\item
Right-handed neutrinos can have normal, dimension-four Yukawa couplings to the
lepton doublet. In $SO(10)$ such couplings are pretty much mandatory,
since they are related by symmetry to those responsible for
charge-$\frac{2}{3}$ quark masses.  In addition, being neutral under
$SU(3)\times SU(2)\times U(1)$
they, unlike the fermions of the standard
model, can have a Majorana type self-mass without breaking these low-energy
symmetries.   We might expect the self-mass to arise where it is first
allowed, at the scale where $SO(10)$  breaks (or its moral equivalent).
Masses of that magnitude remove these particles from the accessible
spectrum, but they have an important indirect effect. In second-order
perturbation theory the ordinary left-handed neutrinos, through their
ordinary Yukawa couplings, make virtual transitions to their right-handed
relatives and back. This generates non-zero masses for the ordinary
neutrinos that are much smaller than the masses of other leptons and quarks.
The magnitudes predicted in this way are broadly consistent with
the observed tiny masses.  No more than order-of-magnitude success
can be claimed, because many relevant details of the models are poorly
determined.
\item
Unification tends to obliterate the distinction between quarks and leptons,
and hence to open up the possibility of proton decay. Heroic experiments
to observe this process have so far come up empty, with limits on partial
lifetimes approaching $10^{34}$ years for some channels. It is very difficult
to assure that these processes are sufficiently suppressed, unless the
unification scale is very large.  Even the high scale indicated by running of
couplings and neutrino masses is barely adequate. Spinning it
positively, experiments to search for proton decay remain a most
important and promising probe into unification physics.
\item
Similarly, it is difficult to avoid the idea that unification, brings in new
connections among the different families.  There are significant
experimental constraints on strangeness-changing neutral currents, lepton
number violation, and other exotic processes that must be suppressed,
and this makes a high scale welcome.
\item
Axion physics requires a high scale of Peccei-Quinn symmetry breaking,
in order to implement weakly coupled, ``invisible" axion models.
\item
With the appearance of this large scale, unification of the strong
and electroweak interactions with gravity becomes much more plausible.
Newton's constant has dimensions of mass$^{2}$, so it runs even
classically. Or, to put it another way, gravity responds to energy-momentum,
so it gets stronger at large energy scales. Nevertheless, because
gravity starts out extremely feeble compared to other interactions
on laboratory scales, it becomes roughly equipotent with them only at
enormously high scales, comparable to the Planck energy
$\sim 10^{18}$ Gev. By inverting this thought, we gain a deep insight
into one of the
main riddles about gravity: If gravity is a primary feature of Nature,
reflecting the basic structure of space-time, why does it ordinarily appear so
feeble?  Elsewhere, I have tracked the answer down to the fact that at the
unification (Planck) scale the strong coupling $g_s$ is about $\frac{1}{2}$!
\end{itemize}

These considerations delineate a compelling research program, centered
on gathering more evidence for, and information about, the unification
of fundamental interactions. We need to find low-energy supersymmetry,
and to look hard for proton decay and for axions.  And we need to be
alert to the possibility of direct information from extreme astrophysical
objects and their relics. Such objects include, of course,
the early universe as a whole, but also perhaps contemporary cosmic
defects (strings, domain walls).
They could leave their mark in microwave background anisotropies and
polarization, in gravity waves, or as sources of unconventional and/or
ultra-high energy cosmic rays.

Theoretical suggestions for enhancing the other two components of our
standard model of fundamental physics are less well formed.   I'll
confine myself to a few brief observations.

Non-minimal coupling terms arise in the extension of supersymmetry
to include gravity. Such terms play an important role in many models of
supersymmetry breaking. Although it will require a lot of detective work
to isolate and characterize such terms, they offer a unique and potentially
rich source of information about the role of gravity in unification.

The flavor/Higgs sector of fundamental physics is its least satisfactory part.
Whether measured by the large number of independent parameters or by
the small number of powerful ideas it contains, our theoretical
description of this sector does not attain the same level as we've reached in
the other sectors.   This part really does deserve to be called a ``model"
rather than a ``theory".   There are many opportunities for experiments to
supply additional information.  These include determining masses, weak
mixing angles and phases for quarks; the same for neutrinos; searches for
$\mu \rightarrow e\gamma$ and allied processes; looking for electric
dipole moments; and others. If low-energy supersymmetry is indeed
discovered, there will be many additional masses and mixings to sort out. The
big question for theorists is: What are we going to do with this information?
We need some good ideas that will relate these hard-won answers to
truly fundamental questions.

\section{From Answers to Questions: Cosmology}

Cosmology has been ``reduced" to some general hypotheses and just
four exogenous parameters.
It is an amazing development. Yet I think that most physicists will
not, and should not,
feel entirely satisfied with it. The parameters appearing in
the cosmological model, unlike those in the comparable models of matter, do not
describe the fundamental behavior of simple entities. Rather they
appear as summary descriptors of averaged properties of macroscopic
(VERY macroscopic!) agglomerations. They appear neither as key players in
a varied repertoire of phenomena nor as essential elements in a beautiful
mathematical theory.  Due to these shortcomings we are left wondering
why just these parameters appear necessary to make a working description of
existing observations, and uncertain whether we'll need to include
more as observations are refined. We'd like to carry the analysis to
another level,
where the four working parameters will give way to different
ones that are closer to fundamentals.

There are many ideas for how an asymmetry between matter and antimatter, which
after much mutual annihilation could boil down to the present baryon
density, might
be generated in the early Universe.  Several of them seem
capable of giving the observed value. Unfortunately the answer
generally depends
on details of particle physics at energies that are unlikely to be
accessible experimentally any time soon. So for a decision among them we may be
reduced to waiting for a functioning Theory of (Nearly)
Everything.

I'm much more optimistic about the dark matter problem.  Here we have
the unusual
situation that there are two good ideas, which according to William of Occam
(of razor fame) is one too many.  The symmetry of the standard
model can be enhanced, and some of its aesthetic shortcomings can be overcome,
if we extend it to a larger theory.  Two proposed extensions,
logically independent of one another, are particularly specific and compelling.
One of these incorporates a symmetry suggested by Roberto Peccei
and Helen Quinn. PQ symmetry rounds out the logical structure of QCD, by
removing QCD's potential to support strong violation of time-reversal
symmetry, which is not observed.  This extension predicts the existence of
a remarkable new kind of very light, feebly interacting particle:  axions.
The other incorporates supersymmetry, an extension of special relativity
to include quantum space-timed transformations. Supersymmetry
serves several important qualitative and quantitative purposes in modern
thinking about unification, relieving difficulties with understanding why
W bosons are as light as they are and why the couplings of the standard
model take the values they do. In many implementations of
supersymmetry the lightest supersymmetric particle, or LSP, interacts
rather feebly with ordinary matter (though much more strongly than do
axions) and is stable on cosmological time scales.

The properties of these particles, axion or LSP, are just right for
dark matter.
Moreover you can calculate how abundantly they would be produced in the
Big Bang, and in both cases the prediction for the abundance is quite
promising.  There are vigorous, heroic experimental searches underway
to dark matter in either of these forms. We will also get crucial
information about supersymmetry from the Large Hadron Collider (LHC),
starting in 2007. I will be disappointed -- and surprised -- if we don't have
a much more personalized portrait of the dark matter in hand a decade from now.

It remains to say a few words about the remaining parameter, the density
of dark energy.  There are two problems with this: Why it is so
small? Why is it so big?

A great lesson of the standard model is that what we have been evolved to
perceive as empty space is in fact a richly structured medium. It contains
symmetry-breaking condensates associated with electroweak
superconductivity and spontaneous chiral symmetry breaking in QCD, an
effervescence of virtual particles, and probably much more. Since
gravity is sensitive to all forms of energy it really ought to see this stuff,
even if we don't.  A straightforward estimation suggests that empty space
should weigh several orders of magnitude of orders of magnitude
(no misprint here!) more than it does.  It ``should" be much denser than a
neutron star, for example. The expected energy of empty space acts like
dark energy, with negative pressure, but there's much too much of it.

To me this discrepancy is the most mysterious fact in all of physical science,
the fact with the greatest potential to rock the foundations.  We're obviously
missing some major insight here. Given this situation, it's hard to know
what to make of the ridiculously small amount of dark energy that
presently dominates the Universe!

The emerging possibility of forging links between fundamental physics and
cosmology through models of inflation is good reason for excitement
and optimism.
Several assumptions in the standard cosmological model,
specifically uniformity, spatial flatness, and the scale invariant, Gaussian,
adiabatic (Harrison-Zeldovich) spectrum, were originally suggested on
grounds of simplicity, expediency, or esthetics.  They can be supplanted
with a single dynamical hypothesis: that very early in its history the
Universe underwent a period of superluminal expansion, or inflation.
Such a period could have occurred while a matter field that was coherently
excited out of its ground state permeated the Universe. Possibilities of
this kind are easy to imagine in models of fundamental physics. For example
scalar fields are used to implement symmetry breaking even in the
standard model, and such fields can easily fail to shed energy
quickly enough to
stay close to their ground state as the Universe expands. Inflation
will occur if the approach to the ground state is slow enough.
Fluctuations will be
generated because the relaxation process is not quite synchronized
across the Universe.

Inflation is a wonderfully attractive, logically compelling idea, but
very basic
challenges remain.  Can we be specific about the cause of inflation,
grounding it
in specific, well-founded, and preferably beautiful models of
fundamental physics?  Concretely, can we calculate the correct amplitude of
fluctuations convincingly?  Existing implementations actually have a
problem here; it takes some nice adjustment to get the amplitude
sufficiently small.

More hopeful, perhaps, than the difficult business of extracting hard
quantitative
predictions from a broadly flexible idea, is to follow up on the
essentially new and
surprising possibilities it suggests.  The violent restructuring of
space-time attending inflation should generate detectable gravitational waves.
These can be detected through their effect on polarization of the
microwave background. And the non-trivial dynamics of relaxation should
generate some detectable deviation from a strictly scale-invariant
spectrum of fluctuations. These are very well posed questions, begging for
   experimental answers.

Perhaps not quite so sharply posed, but still very promising, is the
problem of the
origin of the highest energy cosmic rays. It remains controversial
whether there
so many events observed at energies above those where protons
or photons could travel cosmological distances that explaining their existence
requires us to invoke new fundamental physics. However this plays
out, we clearly have a lot to learn about the compositions of these
events, their sources,
and the acceleration mechanisms. \vspace{-7pt}

\section{Is the Universe a Strange \emph{Place}?}
The observed values of the ratios $\rho_\Lambda/\rho_{\rm DM}$ and
$\rho_{\rm DM}/\rho_b$ are extremely peculiar from the point of view
of fundamental physics, as currently understood. Leading ideas from
fundamental theory about the origin of dark matter and the origin of
baryon number ascribe them to causes that are at best very remotely
connected, and existing physical ideas about the dark energy, which are
sketchy at best, don't connect it to either of the others. Yet the ratios are
observed to be close to unity. And the fact that these ratios are close to
unity is crucial to cosmic ecology; the world would be a very different place
if their values were grossly different from what they are.

Several physicists, among whom S. Weinberg was one of the earliest and
remains among the most serious and persistent, have been led to wonder
whether it might be useful, or even necessary, to take a different
approach, invoking anthropic reasoning. Many physicists view such
reasoning as a compromise or even a betrayal of the goal of understanding the
world in rational, scientific terms. Certainly, some adherents of the
``Anthropic Principle" have overdone it. No such ``Principle'' can
substitute for deep principles like
symmetry and locality, which support a vast wealth of practical and
theoretical applications, or the algorithmic description of Nature in
general. But
I believe there are specific, limited circumstances in which anthropic
reasoning is manifestly appropriate and unavoidable.

In fact, I will now sketch an existence proof.

\subsection{Relevant Properties of Axions}

I will need to use a few properties of axions, which I should briefly
recall.

Given its extensive symmetry and the tight structure of relativistic quantum
field theory, the definition of QCD only requires, and only permits, a very
restricted set of parameters. These consist of the coupling constant and
the quark masses, which we've already discussed, and one more -- the so-called
$\theta$~parameter.  Physical results depend periodically upon
$\theta$, so that effectively it can take values between $\pm \pi$.
We don't know the actual value of the
$\theta$ parameter, but only a limit, $|\theta | \sim 10^{-9}$.
Values outside this small range are excluded by experimental results,
principally the tight bound on the electric dipole moment of the neutron. The
discrete symmetries P and T are violated unless $\theta \equiv 0$ (mod $~\pi$).
Since there are P and T violating interactions in the world, the
$\theta$ parameter can't be to zero by any strict symmetry assumption.
So understanding its smallness is a challenge.

The effective value of $\theta$ will be affected by dynamics, and in particular
by spontaneous symmetry breaking. Peccei and Quinn discovered that if one
imposed a certain asymptotic symmetry, and if that symmetry were
broken spontaneously, then an effective value $\theta \approx 0$
would be obtained.
Weinberg and I explained that the approach
$\theta \rightarrow 0$ could be understood as a relaxation process,
whereby a very
light field, corresponding quite directly to $\theta$, settles into
its minimum energy state.
This is the axion field, and its quanta are called axions.

The phenomenology of axions is essentially controlled by one
parameter, $F$.  $F$ has
dimensions of mass. It is the scale at which Peccei-Quinn symmetry breaks.

\subsection{Cosmology}

Now let us consider the cosmological implications.
Peccei-Quinn symmetry is unbroken at temperatures $T\gg F$.
When this symmetry breaks the initial value of the phase, that is
$e^{ia/F}$, is random beyond the then-current horizon scale. One can
analyze the fate of these fluctuations by solving the equations for a
scalar field in an expanding Universe.

The main general results are as follows.  There is an effective cosmic
viscosity, which
keeps the field frozen so long as the Hubble parameter
$H \equiv \dot{R} /R \gg  m$, where $R$ is the expansion factor.
   In the opposite limit
$H \ll m$ the field undergoes lightly damped oscillations, which result in
an energy density that decays as $\rho \propto 1/R^3$.  Which is to say,
a comoving volume contains a fixed mass. The field can be regarded as a
gas of nonrelativistic particles (in a coherent state). There is some
additional
damping at intermediate stages. Roughly speaking we may say that
the axion field, or any scalar field in a classical regime, behaves
as an effective
cosmological term for $H>>m$ and as cold dark matter for $H\ll m$.
Inhomogeneous perturbations are frozen in while their length-scale exceeds
$1/H$, the scale of the apparent horizon, then get damped.

If we ignore the possibility of inflation, then there is a unique
result for the cosmic
axion density, given the microscopic model. The criterion $H \sim m$
is satisfied for
$T\sim \sqrt {M_{\rm Planck}/ F}
\Lambda_{\rm QCD}$. At this point the horizon-volume contains many
horizon-volumes
from the Peccei-Quinn scale, but it still contains only a negligible
amount of energy
by contemporary cosmological standards. Thus in comparing to
current observations, it is appropriate to average over the starting
amplitude $a/F$ statistically.
If we don't fix the baryon-to-photon ratio, but
instead demand spatial flatness, as inflation suggests we should,
then for $F > 10^{12}$ Gev the
baryon density we compute is smaller than what we
observe.

If inflation occurs before the Peccei-Quinn transition, this analysis
remains valid.
But if inflation occurs after the transition, things are quite different.

\subsection{Anthropic Reasoning Justified}

For if inflation occurs after the transition, then the
patches where $a$ is approximately homogeneous get
magnified to enormous size. Each one is far larger than the presently
observable Universe.
The observable Universe no longer contains a fair statistical sample of
$a/F$, but some particular ``accidenta" value.  Of course there
is still a larger structure, which Martin Rees calls the Multiverse,
over which the value varies.

Now if $F>10^{12}$ Gev, we could still be consistent with
cosmological constraints on the axion density, so long as
the amplitude satisfies
$ (a/F )^2 \sim F/( 10^{12}~{\rm Gev})$.  The actual value of $a/F$,
which controls
a crucial regularity of the observable Universe, is contingent in a very
strong sense. In fact, it is different ``elsewhere".

Within this scenario, the anthropic principle is demonstrably correct
and appropriate.
Regions having large values of $a/F$, in which axions by far dominate
baryons, seem likely to prove inhospitable for the development of complex
structures. Axions themselves are weakly interacting and essentially
dissipationless, and they dilute the baryons, so that these too stay dispersed.
In principle laboratory experiments could discover axions with $F >
10^{12}$ Gev.
If they did, we would have to conclude that the vast bulk of the
Multiverse was inhospitable to intelligent life. And we'd be forced to appeal
to the anthropic principle to understand the anomalously modest axion
density in our Universe.

Weinberg considered anthropic reasoning in connection with the density
of dark energy. It would be entertaining to let both densities, and perhaps
other parameters, float simultaneously, to see whether anthropic
reasoning favors the observed values. Of course, in the absence of a
plausible microscopic setting, these are quite speculative exercises. We don't
know, at present, which if any combinations of the basic parameters
that appear in our desciption of Nature vary independently over the
Multiverse.  But to the extent anthropic reasoning succeeds, it might
guide us toward some specific hypotheses about fundamental physics (e.g.,
that axions provide the dark matter, that $F>10^{12}$ Gev, that the
dark matter candidates suggested by supersymmetry are subdominant, or
perhaps unstable on cosmological time scales).

One last thought, inspired by these considerations. The essence of
the Peccei-Quinn mechanism is to promote the phase of quark mass
matrix to an independent, dynamically variable field. Could additional aspects
of the quark and lepton mass matrices likewise be represented as
dynamical fields?  In fact, this sort of set-up appears quite naturally in
supersymmetric models, under the rubric ``flat directions" or ``moduli".
Under certain not entirely implausible conditions particles associated
with these moduli fields could be accessible at future accelerators,
specifically the LHC. If so, their study could shed new light on the
family/Higgs
sector, where we need it badly.

\section{Convergence}

The way in which many of our most ambitious questions, arising from
the perimeters of logically independent circles of ideas, overlap and link
up is remarkable. It might be a sign that we poised to break through to a
new level of integration in our understanding of the physical world. Of course,
to achieve that we will need not only sharp ambitious questions, but
also some convincing answers. There are many promising lines to pursue,
as even this brief and very incomplete discussion has revealed.

\end{document}